\input harvmac.tex

%%%%%%%%%%%%%%%%%%%%% Rublenye bukvy%%%%%%%%%%%%%%%%%%%%%
\def\IB{\relax\hbox{$\inbar\kern-.3em{\rm B}$}}
\def\IC{\relax\hbox{$\inbar\kern-.3em{\rm C}$}}
\def\ID{\relax\hbox{$\inbar\kern-.3em{\rm D}$}}
\def\IE{\relax\hbox{$\inbar\kern-.3em{\rm E}$}}
\def\IF{\relax\hbox{$\inbar\kern-.3em{\rm F}$}}
\def\IG{\relax\hbox{$\inbar\kern-.3em{\rm G}$}}
\def\IGa{\relax\hbox{${\rm I}\kern-.18em\Gamma$}}
\def\IH{\relax{\rm I\kern-.18em H}}
\def\IK{\relax{\rm I\kern-.18em K}}
\def\II{\relax{\rm I\kern-.18em I}}
\def\IL{\relax{\rm I\kern-.18em L}}
\def\IP{\relax{\rm I\kern-.18em P}}
\def\IR{\relax{\rm I\kern-.18em R}}
\def\IZ{\relax\ifmmode\mathchoice
{\hbox{\cmss Z\kern-.4em Z}}{\hbox{\cmss Z\kern-.4em Z}}
{\lower.9pt\hbox{\cmsss Z\kern-.4em Z}}
{\lower1.2pt\hbox{\cmsss Z\kern-.4em Z}}\else{\cmss Z\kern-.4em Z}\fi}

%%%%%%%%%%%%%%%%%%%% Calligraphic letters%%%%%%%%%%%%%%%%%%%
\def\CA {{\cal A}}

\def\CD {{\cal D}}
\def\CE {{\cal E}}
\def\CF {{\cal F}}

\def\CM {{\cal M}}
\def\CN {{\cal N}}

\def\CS {{\cal S}}

\def\CW {{\cal W}}

%%%%%%%%%%%%%%% Cyrillic %%%%%%%%%%%%%

\chardef\tempcat=\the\catcode`\@
\catcode`\@=11
\def\cyracc{\def\u##1{\if \i##1\accent"24 i%
\else \accent"24 ##1\fi }}
\newfam\cyrfam
\font\tencyr=wncyr10
\def\cyr{\fam\cyrfam\tencyr\cyracc}

%%%%%%%%%%%%%%%%%% Derivatives %%%%%%%%%%%%%%%%%%%%%%%%
\def\p{\partial}
\def\pa{\partial}

%%Beltrami

\def\be{\mu^z_\zb}
\def\beb{\mu^\z_z}

%%%%%%%%%%%%%%%% letters with bar %%%%%%%%%%%%%%%%%%%%%%%

\def\zb {{\bar{z}}}
\def\z{{\bar {z}}}      
\def\ab{{\bar a}}

\def\LB{{\bar L}}
%%%%%%%%%%%%%%%% Math symbols %%%%%%%%%%%%%%%%%%%%%%%

\def\Tr{{\rm Tr}}

%%%%%%%%%%%%%%%%%%%%% Short Cuts %%%%%%%%%%%%%%%%%%%%%%%

\def\demi{{1\over 2}}
\def\half {{1\over 2}}

\def\ymc{g_{{}_{\rm YM}}}
\def\gs{g_{{}_{\rm S}}}

%%%%%%%%%%%%%%%%%% Greek %%%%%%%%%%%%%%%%%%%%%%

\def\f{\phi} \def\fb{{\bar \f}}
\def\P{\Psi} 
\def\F{\Phi} 
\def\vp{\varphi}\def\vpb{{\bar \varphi}}

\def\a{\alpha}
\def\b{\beta}
\def\g{\gamma} 
 
\def\m{\mu}

\def\ve{\varepsilon}

\def\l{\lambda} \def\L{\Lambda}

\def\Si{{\Sigma}}

%%%%%%%%%%%%%%%%%% Big ( ) %%%%%%%%%%%%%%%%%%%%%%
\def\|{\Big|}
\def\({\Big(} \def\){\Big)}
\def\[{\Big[} \def\]{\Big]}

%%%%%%%%%%%%%%%%%% Text %%%%%%%%%%%%%%%%%%%%%%

\def\ie{{\it i.e.}}
\def\cf{{\it cf.}}

%%%%%%%%%%%%% References %%%%%%%%%%%%%%%%%%%%

\def\paper#1#2#3#4{#1, {\sl #2}, #3 {\tt #4}}
% refs with #1=authors, #2=title, #3=publ.ref, #4=hep no :
%\lref\NAME{\paper
%{Authors}{Title(in \it)}{\PLB{No.}{Year}{page},}
%{\hh 9907099 (in\tt)}.}

%\def\hh#1{hep-th/{\it #1}}
\def\hh{hep-th/}

% journal~{\bf no.} (year) page

\def\PLB#1#2#3{Phys. Lett.~{\bf B#1} (#2) #3}
\def\NPB#1#2#3{Nucl. Phys.~{\bf B#1} (#2) #3}
\def\PRL#1#2#3{Phys. Rev. Lett.~{\bf #1} (#2) #3}
\def\CMP#1#2#3{Comm. Math. Phys.~{\bf #1} (#2) #3}
\def\PRD#1#2#3{Phys. Rev.~{\bf D#1} (#2) #3}
\def\MPL#1#2#3{Mod. Phys. Lett.~{\bf #1} (#2) #3}
\def\IJMP#1#2#3{Int. Jour. Mod. Phys.~{\bf #1} (#2) #3}

%%%%%%%%% Something to deal with sub-sub-sections%%%%%%%%%

\def\unlockat{\catcode`\@=11}

\unlockat

% Something to deal with sub-sub-sections

\def\newsec#1{\global\advance\secno by1\message{(\the\secno. #1)}
\global\subsecno=0\global\subsubsecno=0\eqnres@t\noindent
{\bf\the\secno. #1}
\writetoca{{\secsym} {#1}}\par\nobreak\medskip\nobreak} %
\global\newcount\subsecno \global\subsecno=0
\def\subsec#1{\global\advance\subsecno
by1\message{(\secsym\the\subsecno. #1)}
\ifnum\lastpenalty>9000\else\bigbreak\fi\global\subsubsecno=0
\noindent{\it\secsym\the\subsecno. #1}
\writetoca{\string\quad {\secsym\the\subsecno.} {#1}}
\par\nobreak\medskip\nobreak}
\global\newcount\subsubsecno \global\subsubsecno=0
\def\subsubsec#1{\global\advance\subsubsecno by1
\message{(\secsym\the\subsecno.\the\subsubsecno. #1)}
\ifnum\lastpenalty>9000\else\bigbreak\fi
\noindent\quad{\secsym\the\subsecno.\the\subsubsecno.}{#1}
\writetoca{\string\qquad{\secsym\the\subsecno.\the\subsubsecno.}{#1}}
\par\nobreak\medskip\nobreak}

\def\subsubseclab#1{\DefWarn#1\xdef
#1{\noexpand\hyperref{}{subsubsection}%
{\secsym\the\subsecno.\the\subsubsecno}%
{\secsym\the\subsecno.\the\subsubsecno}}%
\writedef{#1\leftbracket#1}\wrlabeL{#1=#1}}% Macros for boxes \lockat

%why???\font\manual=manfnt
\def\dbend{\lower3.5pt\hbox{\manual\char127}}

\def\sdtimes{\mathbin{\hbox{\hskip2pt\vrule
height 4.1pt depth -.3pt width .25pt \hskip-2pt$\times$}}}

%%%%%%%%%%%%%%%%%%% Macros for boxes %%%%%%%%%%%%%%%%%%

\def\boxit#1{\vbox{\hrule\hbox{\vrule\kern8pt
\vbox{\hbox{\kern8pt}\hbox{\vbox{#1}}\hbox{\kern8pt}}
\kern8pt\vrule}\hrule}}

\def\mathboxit#1{\vbox{\hrule\hbox{\vrule\kern8pt\vbox{\kern8pt
\hbox{$\displaystyle #1$}\kern8pt}\kern8pt\vrule}\hrule}}

%%%%%%%%%%%%%%%% ANOTHER SET OF MACROS %%%%%%%%%%%%%%%%%%

\def\inbar{\,\vrule height1.5ex width.4pt depth0pt}

\font\cmss=cmss10 \font\cmsss=cmss10 at 7pt

\def\sdtimes{\mathbin{\hbox{\hskip2pt\vrule
height 4.1pt depth -.3pt width .25pt\hskip-2pt$\times$}}}

%REFERENCES
%%%%%%%%%%%%%%%%%%%%%%%%%%%%%%%%%%%%%%%%%%%%%%%%%%%

\lref\simons{
J. Cheeger and J. Simons, {\it Differential Characters
and Geometric Invariants}, Stony Brook Preprint, (1973),
unpublished.}

\lref\cargese{
L.~Baulieu, {\it Algebraic quantization of gauge theories}, Perspectives 
in
fields and particles, Plenum Press, eds. Basdevant-Levy,
Cargese Lectures 1983}

\lref\antifields{
L. Baulieu, M. Bellon, S. Ouvry, C.Wallet, Phys.Letters B252
(1990) 387; M. Bocchichio,
Phys. Lett. B187 (1987) 322; Phys. Lett. B 192 (1987) 31;
R. Thorn Nucl. Phys. B257 (1987) 61. }

\lref\thompson{
George Thompson, Annals Phys. 205 (1991) 130; J.M.F. Labastida,
M. Pernici, Phys. Lett. 212B (1988) 56;
D. Birmingham, M.Blau, M. Rakowski and G.Thompson, Phys. Rept.
209 (1991) 129.}

\lref\wittensix{
E. Witten, {\it New Gauge Theories In Six Dimensions}, \hh{9710065}. }

\lref\orlando{
O. Alvarez, L. A. Ferreira and J. Sanchez Guillen, {\it A New Approach to

Integrable Theories in any Dimension}, hep-th/9710147.}

\lref\wittentopo{
E. Witten, {\it Topological Quantum Field Theory}, \hh9403195, Commun.
Math. Phys. {117} (1988)353. }

\lref\wittentwist{
E. Witten, {\it Supersymmetric Yang--Mills theory on a four-manifold},
J. Math. Phys. {35} (1994) 5101.}

\lref\west{
L.~Baulieu, P.~West, {\it Six Dimensional TQFTs and Self-dual 
Two-Forms.}}

\lref\bv{
I.A. Batalin and V.A. Vilkowisky, Phys. Rev. D28 (1983) 2567\semi M.
Henneaux, Phys. Rep. 126 (1985) 1\semi
M. Henneaux and C. Teitelboim, {\it
Quantization of Gauge Systems}  Princeton University Press,
Princeton (1992).}

\lref\kyoto{
L.~Baulieu, E. Bergschoeff and E. Sezgin, Nucl. Phys. {\bf B}307 (1988)
348\semi
L.~Baulieu, ``Field Antifield Duality, p-Form Gauge Fields
and Topological Quantum Field Theories'', hep-th/9512026, Nucl. Phys.
{\bf B} 478 (1996) 431. }

\lref\sourlas{
G. Parisi and N. Sourlas, ``Random
Magnetic Fields, Supersymmetry and Negative Dimensions'', Phys. Rev.
Lett. 43 (1979) 744; Nucl. Phys. {\bf B} 206 (1982) 321. }

\lref\SalamSezgin{
A. Salam and E. Sezgin,
{\it Supergravities in diverse dimensions}, vol. 1, p. 119\semi P. Howe,
G. Sierra and P. Townsend, Nucl Phys {\bf B}221 (1983) 331.}

%%%%%%
%% References herein
%%%%%%%%%%%%%%%%%%%%%%%%%%%%%%%%%%%%%%%%%%%%%%%%

\def\ll{\vskip 1.5mm}

%%\lref\NAME{\paper
%%{Authors}{Title(in \sl)}{\PLB{No.}{Year}{page},}
%%{\hh 9907099 (in\tt)}.}

%%% 7sep99 refs. added -> version to be published in PLB

\lref\periw{\paper
{V.~Periwal}{The Topology of Events}{}{\hh 9803150.}}

\lref\mnmz{\paper
{A.~Losev, G.~Moore, 
S.~Shatashvili}{M \& m's}{\NPB{522}{1998}{105},}{\hh 9707250.}}

\lref\sugino{
{F.~Sugino,}{\sl Cohomological Field Theory
Approach to Matrix Strings,} {\tt \hh 9904122.}}

\lref\topmat{\paper
{S.~Hirano, M.~Kato}{Topological Matrix Model}{Prog. Theor. Phys. {\bf 
98} (1997)
1371}{\hh 9708039.}}

\lref\Berkovits{\paper
{N.~Berkovits}{Quantization of the Superstring with Manifest U(5)
Super-Poincare Invariance}{}{\hh 9902099.}}

\lref\nicolai{\paper
{H.~Nicolai}{New Linear Systems for 2D Poincar\'e
Supergravities}{\NPB{414}{1994}{299},}{\hh 9309052.}\ll}

\lref\higher{\paper{L.~Baulieu, A.~Losev, N.~Nekrasov}{Chern-Simons
and Twisted Supersymmetry in Higher
Dimensions}{\NPB{522}{1998}{82},}{hep-th/9707174.}\ll}

\lref\joelight{\paper
{J.~Polchinski}{M-Theory and the Light-Cone}{}{\hh 9903165}.\ll}

\lref\dima{\paper
{V.~Knizhnik}{Analytic fields on Riemann
surfaces}{\PLB{180}{1986}{247}.}{}\ll}

\lref\narns{\paper
{H.S.~Yang, I.~Kim, B.H.~Lee}{Non-abelian Ramond-Neveu-Schwarz String
Theory}{}{\hh 9802176.}\ll}

\lref\ohta{\paper
{L.~Baulieu, N.~Ohta}{Worldsheets with Extended
Supersymmetry} {\PLB{391}{1997}{295},}{\hh 9609207.}\ll}

\lref\bg{\paper
{L.~Baulieu, B.~Grossman}{Monopoles and Topological
Field Theory}{\PLB{214}{1988}{223}.}{}\ll}

\lref\little{\paper
{M.~Berkooz, M.~Rozali, N.~Seiberg}{Matrix Description of M--theory on T4
and T5}{\PLB{408}{1997}{105},}{\hh 9704089}\semi
\paper
{N.~Seiberg}{Matrix Description of M--theory on T5 and T5/Z2}{\PLB{408}
{1997}{98},}{\hh 9705221.}
 }

\lref\seibergsix{\paper
{N.~Seiberg}{Non-trivial Fixed Points of The Renormalization
Group in Six
Dimensions}{\PLB{390}{1997}{169}}{\hh 9609161}\semi \paper
{O.J.~Ganor, D.R.~Morrison, N.~Seiberg}
{Branes, Calabi-Yau Spaces, and Toroidal Compactification of the N=1
Six-Dimensional $E_8$ Theory}{\NPB{487}{1997}{93}}{\hh 9610251}\semi
\paper
 {O.~Aharony, M.~Berkooz, N.~Seiberg}{Light-Cone
Description of (2,0) Superconformal Theories in Six
Dimensions}{Adv. Theor. Math. Phys.
{\bf 2} (1998) 119}{\hh 9712117.}\ll}

\lref\cs{\paper
{L.~Baulieu}{Chern-Simons Three-Dimensional and Yang-Mills-Higgs
Two-Dimensional Systems as
Four-Dimensional Topological Quantum Field
Theories}{\PLB{232}{1989}{473}.}{}\ll}

\lref\beltrami{\paper
{L.~Baulieu, M.~Bellon}{Beltrami Parametrization and String
Theory}{\PLB{196}{1987}{142}}{}\semi
\paper
{L.~Baulieu, M.~Bellon, R.~Grimm}{Beltrami Parametrization For
Superstrings}{\PLB{198}{1987}{343}}{}\semi
\paper
{R.~Grimm}{Left-Right Decomposition of Two-Dimensional
Superspace Geometry
and Its BRS Structure}{Annals Phys. {\bf 200} (1990) 49.}{}\ll}

\lref\bbg{\paper
{L.~Baulieu, M.~Bellon, R.~Grimm}{Left-Right Asymmetric
Conformal Anomalies}{\PLB{228}{1989}{325}.}{}\ll}

\lref\bonora{\paper
{G.~Bonelli, L.~Bonora, F.~Nesti}{String Interactions from
Matrix String Theory}{\NPB{538}{1999}{100},}{\hh 9807232}\semi
\paper
{G.~Bonelli, L.~Bonora, F.~Nesti, A.~Tomasiello}{Matrix String
Theory and its Moduli Space}{}{\hh 9901093.}\ll}

\lref\corrigan{\paper
{E.~Corrigan, C.~Devchand, D.B.~Fairlie, J.~Nuyts}{First Order
Equations for Gauge Fields in Spaces of Dimension
Greater Than
Four}{\NPB{214}{452}{1983}.}{}\ll}

\lref\acha{\paper
{B.S.~Acharya, M.~O'Loughlin, B.~Spence}{Higher Dimensional
Analogues of Donaldson--Witten
Theory}{\NPB{503}{1997}{657},}{\hh 9705138}\semi
\paper {B.S.~Acharya, J.M.~Figueroa-O'Farrill, M.~O'Loughlin,
B.~Spence}{Euclidean
D-branes and Higher-Dimensional Gauge
Theory}{\NPB{514}{1998}{583},}{\hh
9707118.}\ll}

\lref\Witr{\paper{E.~Witten}{Introduction to Cohomological Field 
Theories}
{Lectures at Workshop on Topological Methods in Physics
(Trieste, Italy, Jun 11-25, 1990), \IJMP{A6}{1991}{2775}.}{}\ll}

\lref\ohta{\paper
{L.~Baulieu, N.~Ohta}{Worldsheets with Extended
Supersymmetry} {\PLB{391}{1997}{295},}{\hh 9609207.}\ll}

\lref\gravity{\paper
{L.~Baulieu}{Transmutation of Pure 2-D
Supergravity Into Topological 2-D Gravity and Other
Conformal Theories} {\PLB{288}{1992}{59},}{\hh 9206019.}\ll}

\lref\wgravity{\paper
{L.~Baulieu, M.~Bellon, R.~Grimm}{Some Remarks on the
Gauging of the Virasoro
and $w_{1+\infty}$ Algebras}{\PLB{260}{1991}{63}.}{}\ll}

\lref\fourd{\paper{E.~Witten}{Topological Quantum Field
Theory}{\CMP{117}{1988}{353}}{}\semi
\paper
{L.~Baulieu, I.M.~Singer}{Topological Yang--Mills
Symmetry}{Nucl. Phys. Proc. Suppl. {\bf 15B} (1988) 12.}{}\ll}

\lref\topo{\paper
{L.~Baulieu}{On the Symmetries of Topological Quantum
Field Theories}{\IJMP{A10}{1995}{4483},}{\hh 9504015}\semi
\paper
{R.~Dijkgraaf, G.~Moore}{Balanced Topological Field
Theories}{\CMP{185}{1997}{411},}{\hh 9608169.}\ll}

\lref\wwgravity{\paper
{I.~Bakas} {The Large $N$ Limit of
Extended Conformal Symmetries}{\PLB{228}{1989}{57}.}{}\ll}

\lref\wwwgravity{\paper
{C.M.~Hull}{Lectures on $\CW$-Gravity, $\CW$-Geometry and
$\CW$-Strings}{}{\hh 9302110}, and~references therein.\ll}

\lref\wvgravity{\paper
{A.~Bilal, V.~Fock, I.~Kogan}{On the origin of
 $\CW$-algebras}{\NPB{359}{1991}{635}.}{}\ll}

\lref\surprises{\paper
{E.~Witten} {Surprises with
Topological Field Theories} {Lectures given at
``Strings 90'', Texas A\&M, 1990,}{Preprint IASSNS-HEP-90/37.}\ll}

\lref\stringsone{\paper
{L.~Baulieu, M.B.~Green, E.~Rabinovici}{A Unifying Topological
Action for Heterotic and Type II Superstring
Theories}{\PLB{386}{1996}{91},}{\hh 9606080} \semi
\paper
 {L.~Baulieu, M.B.~Green, E.~Rabinovici}{Superstrings from Theories
with $N>1$
World Sheet Supersymmetry} {\NPB{498}{1997}{119},}{\hh 9611136.}\ll}

\lref\bks{\paper
{L.~Baulieu, H.~Kanno, I.~Singer}{Special Quantum Field Theories in
Eight and Other
Dimensions}{\CMP{194}{1998}{149},}{\hh 9704167}\semi
\paper
{L.~Baulieu, H.~Kanno, I.~Singer}{Cohomological Yang--Mills
Theory
in Eight Dimensions}{
Talk given at APCTP Winter School on Dualities in String
Theory (Sokcho, Korea, February 24-28, 1997),} {\hh 9705127.}\ll}

\lref\witdyn{\paper{P.~Townsend}{The Eleven Dimensional Supermembrane
Revisited}{\PLB{350}{1995}{184},}{\hh9501068}\semi
\paper{E.~Witten}{String Theory Dynamics in Various
Dimensions}{\NPB{443}{1995}{85},}{\hh 9503124}.\ll}

\lref\bfss{\paper
{T.~Banks, W.Fischler, S.H.~Shenker, L.~Susskind}{$M$-Theory as a
Matrix Model:
A~Conjecture}{\PRD{55}{1997}{5112},} {\hh9610043.}\ll}

\lref\seibergsen{\paper
{N.~Seiberg}{Why is the Matrix Model Correct?}{\PRL{79}{1997}{3577},} 
{\hh
9710009}\semi
\paper
{A.~Sen}{$D0$ Branes on $T^n$ and Matrix Theory}{Adv. Theor. Math.
Phys.~{\bf 2} (1998) 51,}
{\hh 9709220.}\ll}

\lref\laroche{\paper
{L.~Baulieu, C.~Laroche}
{On Generalized Self-Duality Equations Towards Supersymmetric
Quantum Field
Theories Of Forms}{\MPL{A13}{1998}{1115},}{\hh 9801014.}\ll}

\lref\bsv{\paper
{M.~Bershadsky, V.~Sadov, C.~Vafa}
{$D$-Branes and Topological Field Theories}{\NPB{463}
 {1996}{420},}{\hh9511222.}\ll}

\lref\vafapuzz{\paper
{C.~Vafa}{Puzzles at Large N}{}{\hh 9804172.}\ll}

\lref\dvv{\paper
{R.~Dijkgraaf, E.~Verlinde, H.~Verlinde} {Matrix String
Theory}{\NPB{500}{1997}{43},} {\hh9703030.}\ll}

 \lref\wynter{\paper
{T.~Wynter}{Gauge Fields and Interactions in Matrix String
 Theory}{\PLB{415}{1997}{349},}{\hh9709029.}\ll}

\lref\kvh{\paper
{I.~Kostov, P.~Vanhove}{Matrix String Partition
Functions}{}{\hh9809130.}\ll}

\lref\ikkt{\paper
{N.~Ishibashi, H.~Kawai, Y.~Kitazawa, A.~Tsuchiya} {A Large
$N$ Reduced Model as Superstring}{\NPB{498} {1997}{467},}{\hh 
9612115.}\ll}

\lref\ss{\paper
{S.~Sethi, M.~Stern}
{$D$-Brane Bound States Redux}{\CMP{194}{1998} {675},}{\hh 9705046.}\ll}

\lref\mns{\paper
{G.~Moore, N.~Nekrasov, S.~Shatashvili}
{$D$-particle Bound States and Generalized Instantons}{} {\hh 
9803265.}\ll}

\lref\bsh{\paper
{L.~Baulieu, S.~Shatashvili}
{Duality from Topological Symmetry}{}
{\hh 9811198.}\ll}

%%%%%%

%%%\draft

%%%%%%%%%

\font\tiny=cmr10 at 10truept
\font\cap=cmcsc10 at 11truept
\def\idt{\hskip 3mm}

\Title{\vbox{\baselineskip 12pt
\hbox{hep-th/9907099}
\hbox{NSF-ITP~99-26}
\hbox{HUTP~98/A098}
\hbox{LPTHE~99-26}
\hbox{ITEP-TH~30/99}}}
{\vbox{Remarks on Covariant Matrix Strings}}
\vskip -0.3cm
{\bf Laurent
Baulieu$^{1,2}$, C\'eline Laroche$^{1,3}$, Nikita Nekrasov$^{1,4,5,6}$}

{\cap baulieu, laroche@lpthe.jussieu.fr, nikita@curie.harvard.edu}

\vskip 7mm

{\baselineskip11pt
\tiny
{\idt$^1$ LPTHE, Universit\'es Paris~VI--Paris~VII,
Bo\^{\i }te 126, 4 place Jussieu, F-75252 Paris Cedex 05, France}

{\idt$^{2}$ TH-Division, CERN, 1211 Gen{\`e}ve 23, Switzerland}

{\idt$^3$ (From Sept.99) International School for Advanced Studies,
Via Beirut 2--4, 34014 Trieste, Italy}

{\idt$^4$ Lyman Laboratory for Physics,
Harvard University, Cambridge MA
02138, USA}

{\idt$^5$ Institute for Theoretical and Experimental Physics,
Moscow, Russia, 117259}

{\idt$^6$ Institute for Theoretical
Physics, UC Santa Barbara, CA 93106, USA}

}

\vskip 1cm

\hrule

\medskip
\vskip 0.7cm
{\noindent
We  address the issue of  the worldsheet and spacetime
covariant formulation for matrix strings. The  problem is solved in the
limit of vanishing string coupling. To go
beyond the ${\gs} = 0$ limit, we propose a
topological quantum field theory as a twisted candidate. Our model 
involves
the generalized octonionic or $SU(4)$ instanton equations defined in 
eight
dimensions for a {\sl supersymmetric} $U(N)$
Yang--Mills field living on a special holonomy manifold.
The question of untwisting this matrix model into an anomaly free theory
enlightens the need for an ``extended" $2d$--gravity sector, that we 
suggest
could be (partially twisted) $\CW$--gravity.

\Date{}

\newsec{Introduction}

String dualities predict
the existence of an eleven-dimensional
quantum theory of gravitation~\witdyn\/. So far not much is known
about this mysterious M--theory.
The first indication
towards its existence is the fact that type~IIA superstring theory
can be interpreted as a compactified version of the eleven-dimensional
supersymmetric theory with all Kaluza--Klein states present in the form 
of
$D0-$\/branes. In a certain energy regime, the $D0-$\/branes themselves 
are
well described by the supersymmetric quantum mechanics (SQM) of $N\times
N$ matrices with $16$ supercharges.

The proposal of \bfss, further clarified by \seibergsen, and more 
recently by
\joelight, was to interpret M--theory compactified on a light-like circle
with $N$ units of momentum along it as the $U(N)$ case of such SQM.
This proposal is extended for compactifications of M--theory on any
tori $T^{d}$ of dimension $0 \leq d \leq 3$, with the SQM replaced by
the supersymmetric Yang--Mills theory (SYM) on the dual torus
${\widetilde{T^{d}}}$ with the same gauge group $U(N)$.

In particular, the case
$d=1$, expected to be equivalent to type~IIA superstrings, produces
the (1+1)\/--dimensional SYM on a cylinder. R. Dijkgraaf, E. and H. 
Verlinde
(DVV) have analyzed this theory in a
strong $\ymc$ coupling regime (weak string coupling $\gs$) and
found that one can view this gauge theory as a perturbation of the 
orbifold
conformal theory with target $( {\IR}^8)^N/{\CS}_{N}$~\dvv\/. The number 
$8$
here is the number of scalar matrices $X^{1}, \ldots, X^{8}$ in the gauge
multiplet. The permutation group ${\CS}_{N}$ is the part of the unbroken 
gauge
group $U(N) \to U(1)^{N} \sdtimes \CS_{N}$. This description of the gauge
theory contains the second quantized Hilbert space of free string theory, 
as
well as the string interactions in the form of the irrelevant 
perturbation
corresponding to taking the Yang--Mills coupling~$\ymc$ large but finite.

We wish to address the question of a covariant formulation for
this theory of matrix strings. Indeed, in~\dvv\/, it is formulated as
 a Green--Schwarz (GS)--like superstring theory, thus lacking
covariance with respect to both worldsheet and spacetime symmetries
if written in terms of the free fields. Is
there a way to define a ``covariant matrix strings theory'', analogously 
to that
for the $U(1)$ superstrings~? In other words, can one formulate a
Neveu--Schwarz--Ramond (NSR) {\sl matrix} superstring, that would exhibit
manifest worldsheet and spacetime invariances~?

In the present paper, we discuss several possible steps towards
an answer to these questions.

First of all, in the limit ${\gs} = 0$ the problem has a simple solution.
Namely, consider $N$ copies of the fields entering the $c=0$ conformal
field theory of the NSR string in the conformal gauge: one has the bosons
$X^{\mu}_{i}$, left and right fermions $\psi_{L,R; i}^{\mu}$,
the $2d$ gravity ghosts systems: $b_{i}, c_{i}, \beta_{i}, \gamma_{i}$,
with $\mu = 0, \ldots, 9$, $i = 1, \ldots, N$. Take as an action
for this theory the sum of $N$ copies of the standard
NSR action. We get $c = N \cdot 0 = 0$ superconformal theory.
Now consider the orbifold of this theory by the action of the symmetric 
group
${\CS}_{N}$ which acts on the index~$i$. The theory  obtained
 has the following properties:
\item{1.}
When considered on a cylinder, it contains the long strings
of \dvv\ in the form of the twisted sectors;
\item{2.}
The theory has manifest ten--dimensional Lorentz covariance, acting on 
the
index $\mu$;
\item{3.}
It has ${\CW}$--like chiral algebra, generated by the stress-energy 
tensor
$$
T = \sum_{i} T_{i}
$$
where $T_{i} = \p X^{\mu}_{i} \p X^{\mu}_{i} + \psi^{\mu}_{L; i} \p
\psi^{\mu}_{L;i} + ghosts\cdots $ 
 and its higher spin analogues (see~\vafapuzz\/): ${\CW}_{l} =
\sum_{i} T_{i}^{l}, \quad l = 1, 2, \ldots, N$. 
This algebra can be used to show
that on the cylinder this theory is equivalent to the GS theory of~\dvv.

\noindent
The problem is to go beyond ${\gs}=0$.
Similarly to the recent attempts of producing
the covariant formulation of superstrings in various
backgrounds \Berkovits\/, we shall base our model on a
certain version of topological strings.
We propose a version
based on the $SU(4)$
instanton equations~\bks,\acha\ dimensionally reduced from
eight to two dimensions.
The topological field theory (TQFT) that we construct out of these
turns out to give a particular case of type $\bf B$ topological strings.
To enforce its $2d$--covariance, the system with eight matrices
obtained in this approach, describing
transverse non-abelian degrees of freedom, is coupled to a 
two--dimensional gravity background (a version of the gauged
Landau-Ginzburg topological system, based on the superpotential
${W} = {1\over 6} {\ve}_{abc} {\Tr} \f_{a} [ \f_{b}, \f_{c}]$ arises).
Away from
the commuting limit ${\gs} = 0$, there is a further complication from the
dimension of the Yang--Mills coupling~$\ymc$. To compensate for it, and
maintain conformal invariance still in the context of TQFT, we introduce
other fields, of the Wess--Zumino type. This model with eight dimensional
target space now serves as a base to our further steps, concerning 
spacetime
covariance.

In the seek for ten--dimensional covariance, we want to approach
the system with ten matrices.
To this end,
we suggest to ``supersymmetrize'' the latter TQFT: To the
dimensionally reduced Yang--Mills field, we add its eight dimensional
supersymmetric partners. In this manner, we end up with the correct 
number
of matrix coordinates for the possible $SO(10)$ covariance, together with
additional fermions. The extra fields will effectively decouple by
compensation with the corresponding topological ghosts; but we believe 
that
they can be used to produce covariant {\sl heterotic} matrix
strings, via a generalization of~\stringsone\/.
 We now have all ten coordinates at hand (plus superpartners), part of
which
describe unphysical degrees of freedom, naturally leading us to the
question of
anomalies.

In our approach ``\`a la topological'', the anomaly issue arises when
one attempts to untwist the
topological symmetry into spacetime supersymmetry. The idea is that
the anomaly induced by untwisting the fields of the matrix strings sector
should be canceled by a ``countertwist'' of the fields from another 
sector,
namely some extension of topological $2d$--gravity~\stringsone.
The latter is
to be chosen so as to contribute with an  anomaly exactly opposite to 
that
stemming from untwisting the matter sector: The $2d$--gravity
ghost-systems of the untwisted theory must compensate for the anomaly
carried by {\sl matrix} matter fields. In the large $\ymc$ coupling 
limit, the
single-valued fields of the matrix strings theory (the so-called long 
strings)
live on an $N$-sheeted Riemann surface~\wynter, with $N$ the dimension of
the Cartan-subalgebra of the Yang--Mills group. This suggests that the
suitable matrix extension of $2d$--gravity could arise from the
$\CW_N$ part
of partially  untwisted topological $\CW_\infty$ gravity, which
we propose as
a candidate to the ``extended'' $2d$--gravity sector.

\newsec{Generalized Instantons and two--dimensional Covariance}
\subsec{Octonionic or $SU(4)$ Instantons}

Our starting point towards the various directions advocated in this
paper shall be the eight-dimensional theories of~\bks\acha, namely 
twisted
versions of the usual eight-dimensional supersymmetric theory for a
Yang--Mills gauge field $\CA$, of curvature $\CF=d\CA + \CA \CA$, living 
on
a special holonomy manifold $\CM^8$. Two cases are at hand, according to 
the
type of manifold  that one considers. In the case of Spin(7) holonomy, 
one
has octonionic instantons arising from the generalized self-duality 
equations
\eqn\oct{\CF_{i8} \pm \demi c_{ijk} \CF_{jk}=0, \qquad {\rm for} \quad
i,j,\cdots = 1, \ldots, 7}
In the case of $\CM^8$ being a Calabi--Yau fourfold
($CY_4$), it is the $SU(4)$ instantons one deals with, obtained from
\eqn\compx{\eqalign{&\CF_{z^A \z^A} =0, \cr &\CF_{z^A z^B} \pm
\epsilon_{{}_{ABCD}} \CF_{\z^C \z^D} =0, \qquad {\rm for}\quad A,B,
\cdots=1, \ldots,4\cr}}
We shall focus on the latter, most appropriate for
dimensional reduction to two dimensions. Viewing the reduction as the
splitting $\CM^{8} = \CM^{6} \times \Sigma$ where $\Sigma$ is 
two--dimensional
and $\CM^6$ taken to be small,  we are led to split the gauge
field $\CA_{z^{A=1,\ldots,4}}$ 
 as $\f_a \sim \CA_{z^{a=1,\ldots,3}}$ and $A_z\sim
\CA_{z^4}$.

Now, the gauge functions (and thus the Lagrangian that one constructs
using
them) obtained by simplistic reduction of~\compx\ from eight to two
dimensions are not covariant in the sense of two--dimensional
reparametrisations. This is obviously due to the non-commutativity of the
Yang--Mills fields. In order to define a meaningful two--dimensional
theory, we propose to couple our instanton equations to a topological 
gravity
background.

\subsec{Two-Dimensional Covariant Equations}

We will consider our fields propagating on the genus one Riemann
surface $\Si$ (complex coordinates $z, \z$) on which we shall require
$2d$--covariance.\foot{This surface $\Sigma$ should not be confused
with the
string's world-sheet. It is just an interpolating surface, as argued
in~\bonora. Otherwise we would face an overcounting problem.}
We introduce the two--dimensional metric $g_{\alpha\beta}$ in the 
Beltrami
parametrisation, which has the advantage to make left-right factorization
manifest in a conformally invariant way \beltrami\/\nicolai\/:
\eqn\bel{\eqalign{
ds^2 = e_z^+e_\z^- (dz +\be d\zb)(d\zb +\beb dz)
%%\equiv e_z^+e_\z^- (1-\be\beb)^2 dZ \, d\Z
}}
Conformal gauge is restored when one puts the Beltrami differentials
 $\be$,
$\beb$ to zero. The symmetries at hand
allow to set the conformal factor to 1
(\ie\/ $e^+_z$, $e^-_\z$ are both set to 1). Enforcing all needed
two--dimensional covariances for the non-abelian part of our equations
requires the additional introduction of a pair of background left and 
right
Wess--Zumino (W--Z) type fields $L$ and $\bar L$, for which we impose 
that
$e^L$ (resp. $e^\LB$) have $(1,0)$ (resp. $(0,1)$) conformal
weight.\foot{As will be argued in the following, the $L$, $\LB$
fields can possibly be made to propagate and play the role of
compensators in case a conformal anomaly arises after untwisting the 
model.}

With this minimal set of additional background fields, we can now
expect to construct a $2d$--covariant TQFT, based on the following
$2d$--consistent dimensional reduction of~\compx\/:
\eqn\cvrn{\eqalign{& \CE \sim
F_{z\zb} + e^{L+\LB} \sum_{a} [ \f^{a},
\f^{\bar a} ]\cr
& \CE^{a}_{\z} \sim D_{\z} \f^{a} + \half e^{\LB }
\varepsilon^{abc} [\f^{\bar b}, \f^{\bar c}]\cr & \CE^{\bar a}_{z}
\sim D_z \f^{\bar a} + \half e^{L}
\varepsilon^{abc} [\f^{b}, \f^{c}]\cr
}}
The  derivatives above are  covariant  with respect to
both the gauge field $A_z, A_{\zb}$ and the $2d$ gravity. Namely,
for the scalars, it
reads as follows (whereas for higher order tensors, it has
additional  terms involving the Christoffel's):
\eqn\mis{\eqalign{
&D_{\z}\, \f^{a}  = [(\p_{\zb} + A_\z) - \be (\p_z + A_z)]\f^{a} }}

\subsec{Constructing the TQFT: Twisted Light--Cone NSR Matrix Strings}

Making use of the above preliminaries, we can now construct the
two--dimensional TQFT, following the paradigm of cohomological field 
theory
and relying on the eight dimensional construction of~\bks. To do so, we
introduce topological ghosts and ghosts of ghosts for all (dimensionally
reduced) components of $\CA$. Also needed are the antighosts and Lagrange
multipliers adapted to the choice of gauge functions. Altogether, we 
shall deal
with the following set of fields and ghosts, all in adjoint 
representation of
the $U(N)$ gauge group:
\eqn\multipl{
\matrix{&&A_{z,\z}\cr & \P_{z,\z}&&\chi \cr \varphi&&H&&
\bar\varphi \cr &&&\eta } \ ; \quad \matrix{&\f^{a}&&\cr
\P^{{a}}&&\chi^a_\z,
\cr & H^{a}_{\z} \cr &}\ ;
\quad \matrix{&\f^{\ab}&&\cr
\P^{{\ab}}&&\chi^\ab_z \cr & H^{\ab}_{z} \cr &}\ }

In each of the above BRST--multiplet, the first line is a classical
field, the second line is its superpartners (\ie\/ the topological ghosts 
and
antighosts), the third line contains the BRST transform $H$ of the 
antighost
Lagrange multipliers: it is used to enforce the topological gauge 
functions.
In the multiplet for
$A_{z,\bar z}$ there are, in addition,
the bosonic ghosts for ghosts that are responsible for the
underlying Yang--Mills symmetries and the associated
fermionic Lagrange multiplier
$\eta$.\foot{All fields are $U(N)$ matrices, and the $\f^{a}$'s are all 
complex.
We shall
alternatively use
the notations $\f^{\bar a} = \bar\f^{a} =
\f^{a  \dagger}$, where bar means complex conjugation.
The fields $\bar\varphi$
and $ \varphi$ have ghost numbers $-2$ and $2$, respectively and would be
more conventionally denoted as
$\F^{-2}=\bar\varphi$ and $\F^{2}=\varphi$.}

The gauge functions imposed to gauge fix the topological symmetry for the
Yang--Mills field are those of~\cvrn.
They are
organized into three complex equations plus a separate real one,
and accordingly for the antighosts $\chi_\z^a$ and $\chi$.
The  gauge functions that one imposes for the topological ghosts 
$\P_{z,\z}$,
$\P^{a,\bar{a}}$ are constructed in a similar manner: they are the eight
dimensional equations of~\bks, reduced to two dimensions and further
improved  to implement $2d$ covariance:
\eqn\cvgh{
\CE_\P \sim D_z \P_\z +
e^{L+\LB} [\f^{a}, \P^{\ab}] + \ldots
}

In a naive formulation, $L$, $\bar L$ and the Beltrami's $\mu_{\zb}^{z}$,
$\mu_{z}^\z$ are set to zero so that the resulting action is of the form
\eqn\miss{\eqalign{
{\cal I} =  \int_{\Si} \sqrt{g} \ {\Tr}\ & s \Big\{ \chi^{a}_{\zb}  
\left(  \CE^{\bar a}_{z} -
\half\  \ymc^{2} H_{z}^{\bar a} \right) +
\chi^{\bar a}_{z}  \left(  \CE^{a}_{\zb} - \half\ \ymc^{2} H^{a}_{\zb}
\right)  \cr
&-\   \half\ \chi \left( \CE - \half \ymc^{2} H \right)
- \half\ \bar c \left( \p^{\mu} A_{\mu} - \half \ymc^{2} b \right) \
+{1\over{2 \ymc^{2}}}\ \vpb \ \CE_\P \Big\}
}}
where the BRST operator~$s$ only acts on the fields of the matter
sector and not on
the  $2d$--gravity background. (The precise form of the operator $s$ on 
our
various fields can be straightforwardly obtained from that of the eight
dimensional operator in~\bks.)
This action can be interpreted as a twisted version of the light-cone DVV
matrix strings, with
manifest invariance under the eight--dimensional rotations, via the
following correspondence~\stringsone\/ for bosons:
\eqn\bostwist{\eqalign{
& \f^a,\, \fb^a,\, \vp,\, \vpb\,
 \ \to \ X^{m=1\cdots 8}, \cr}}
and fermions:
\eqn\fertwist{\eqalign{
& \P^a,\, \chi^\ab,\, \eta + \chi,\, \P_z \, \ \to \
\psi_-^m, \cr
& \P_\ab,\, \chi_a,\, \eta - \chi,\, \P_\z\, \to \
\psi_+^m . \cr
 }}

\newsec{Spacetime Covariance}

 \subsec{Supersymmetrization and ten--dimensional NSR Fields}

Clearly, the field content of the construction above
cannot allow  invariance under the ten--dimensional group of
rotations. The bosonic sector is based on eightuples of
matrices whereas ten-dimensional invariance  would imply two
more matrices,
describing the longitudinal degrees of freedom.
We propose to get these by replacing
the original set of fields ---the dimensionally reduced field
content of the eight dimensional
Yang--Mills TQFT---
by that of the {\it supersymmetrization} of the same theory.

This means that we shall construct the TQFT for the set of
fundamental fields enlarged to the degrees of freedom of the  full
$\CN=(1,1)$,  $d=8$
 Yang--Mills supermultiplet $( \CA^\m, \l^\a, \f^{\pm})$, where 
$\m=1\cdots
8$ and $ \a=1\cdots 16$.
All fields  are still $U(N)$ matrices, and $ \f^+ =  (\f^-)^{\dag}$.
Building the TQFT for these additional pieces implies the following
additional ``topological partners'':
\eqn\partmult{
 \matrix{&\f^\pm\cr
\P^{\pm}&&\chi^{\pm} \cr&H^{\pm} \cr &}\   ; \quad \matrix{&\l^\a&&\cr
\L^\a&&\chi_\l \cr &H_\l&\cr&}
}
We now  have
ten bosonic matrices to play with: $ X^{M=1\cdots 10} \leftarrow \{ 
X^{m},
\f^\pm \}$, as well as ten left moving fermions $\psi^M_- \leftarrow
\{\psi^m_-, \P^+, \chi^- \}$ and as many right movers $\psi^M_+ 
\leftarrow
\{\psi^m_+, \P^-, \chi^+ \}$.
Another gain is that we end up with  extra fermions from $\l^\a$, which 
can either be decoupled or, more interestingly, be
used to produce a {\sl heterotic}  matrix string in the same spirit
as that of~\stringsone\/.

        \subsec{Does the ``supersymmetrized'' TQFT exhibit manifest
$SO(10)$ invariance~?}

The additional part of the TQFT for the supersymmetric partners is
constructed  as
before, namely by gauging their
topological symmetry.

For the twisted gaugino $\l^\a$, topological freedom and the  requirement
of quantum propagation imply that we introduce a chirality choice; we
choose the right moving condition:
\eqn\gfferm{
\CD_z \l^\a}
with $\CD$ the dimensionally reduced eight dimensional Dirac operator.

 For the gauge functions of the scalars $\f^\pm$, we choose
holomorphicity
condition:
\eqn\gfscal{\eqalign{
D_\z \f^{+}\ \ \  \, ,& \, \ \ D_z \f^{-} .
}}

Note that it
follows from the discussions in~\laroche\/ that there exists no 
ten--dimensional analogue of
 the TQFT of~\bks\acha\/. We could therefore have expected, already from
the eight dimensional point of view, that
 imposing  gauge functions for the $\f^{\pm}$\/'s  would make it 
unavoidable
 to define  the supersymmetrized TQFT on a two-dimensional manifold.

In the $U(1)$ case, our set of fields can easily be interpreted
(with the scheme indicated above) as
ten-dimensional twisted NSR or heterotic superstring fields, in the same manner
as in~\stringsone\/.
However, the question arises whether this remains true in the 
non-commuting
case.
We have expanded the action defined by the above gauge fixings.
Unfortunately, examining the bosonic part shows that the commutator terms
break the $SO(10)$ invariance down to $SO(8) \times SO(2)$.
We haven't yet found a possible modification of our gauge functions that
would enable $SO(10)$ invariance, but we suspect that there might exist
deep reasons for our difficulties besides the mere
lack of imagination.

\newsec{Topological Gravity Side and Anomaly Cancellation}

In the seek of a covariant description, arises the necessity
of having some gauge symmetry allowing to eliminate the
 longitudinal
components  $\f^{\pm}$ and their supersymmetric partners.
 In the ordinary
superstring, this symmetry
is the worldsheet  reparametrisation and local supersymmetry 
invariance.
 The unphysical
coordinates are canceled by the corresponding $(b,c)$, $(\b, \g)$ ghost
systems (more subtle compensations occur  for 
 the construction of the heterotic string). Here,
we need 
the symmetry with functional dimension $2 (N^2-1)$ for it
to be capable of eliminating the two longitudinal $U(N)$ matrices 
$\f^{\pm}$, and
 its generalization for  supersymmetric partners.
The natural choice would be some ``matrix-valued diffeomorphisms'', if
one can make sense of that.

        \subsec{Need of some ``Extended'' two--dimensional Gravity}

We can reformulate this from the ``twisted point of view'':
Our twisted matrix string theory being topological by construction,
has no  local anomaly. Nor can its coupling
to $2d$--gravity  introduce any anomaly: The
 $2d$--gravity must also be of a topological type.
The aim then, is  to construct a theory which,  after untwisting, has the
Ward identities of a superconformal theory.
We are therefore lead to couple our gauge functions to a matrix
extension of the topological $2d$--gravity used for $U(1)$ strings, which 
can
hopefully be untwisted in a way such to compensate for the anomaly 
induced by
untwisting the
 topological matrix string side.

Notice that the Beltrami and {W--Z} fields occurring in~\miss\/ are now 
to
be understood as
background values obtained by
{ gauge-fixing the quantum  fields} of this topological
$2d$--gravity.  Therefore, having  matrix ghosts is in no contradiction
with an
apparently simple $2d$--gravity background: gauge fixing is
purely local.
We have not attempted to understand the moduli which should be introduced
in this theory, and leave unstudied the global aspects.

%%%%%%%%%%%%%%%%%%%%%%%%%%%%%%%%%%%%%

\subsec{Matrix Extension and Anomaly Cancellation}

 Let us now discuss the anomaly question in more details. Our model being
topological in nature, it has a
vanishing conformal anomaly when all fields have the spins
described above, prior to twist.
It is crucial to understand whether it is
justified to  map our twisted fields to physical ones, where
matrices $\f^{M}$ have weight zero, fermions have spin
$\half$ and so on.
Indeed, in the course of doing so, we expect to
generate a super-conformal anomaly from the set of $U(N)$ matter fields.
In the ordinary string case, the anomaly from the matter sector is
compensated by the anomaly from the $(b$, $c)$ and $(\b$, $\g)$
systems of
the  twisted gravity side.
Here, the matter anomaly is that of  $U(N)$ matrix fields, and an 
extended
gravity multiplet is needed to cancel this anomaly.

In the $IR$ limit, the proper (single-valued) matrix string fields,
 interpreted as those for long strings, are defined on an $N$--sheeted
Riemann surface~\dvv\/\wynter.
We suggest to introduce some $N\times N$ matrix $2d$--gravity, with
the diagonal $(e_{z,\z}^\pm)_{ii}$ of the $N\times N$ zweibein describing
gravity on the
$i$'th sheet, whereas the off-diagonal $(e_{z,\z}^\pm)_{ij}$ would
specify
how the $i$'th and $j$'th sheets ``feel'' each other.
Similarly for the associated
 $(b$, $c)$  and $(\b$, $\g)$  matrix ghosts, corresponding to
 having the symmetry of  $N$ independent reparametrisations for each
 sheet.
The ordinary string has,
as a gauge symmetry, the (super)diffeomorphisms ${\rm Diff}(\Sigma)$
times conformal rescalings of the world-sheet metric $h_{\a\b}$.
In particular the first quantized string amplitude is of the form
$$
Z_{1} \sim {1\over{{\rm Diff}(\Sigma)}} \int [\CD X] [\CD h_{\a\b}]\quad
e^{- S(X,h)}.
$$
In a second-quantized picture we
expect amplitudes like:
$$
{{(Z_1)^N}\over{N!}} \sim {1\over{{\rm Diff} ( \Sigma \amalg \ldots
\amalg \Sigma) }} \int [\CD X_{1}] \ldots [\CD X_{N}] [\CD
{h_{\a\b}}^{1}]
\ldots [\CD {h_{\a\b}}^{N}] \quad
e^{- \sum_{i} S(X_{i}, h^{i})}
$$
The group of diffeomorphisms ${\rm Diff}( \Sigma \amalg \ldots
\amalg \Sigma)$ contains, in addition to the $N$ independent
reparametrisations, the permutations which
produce the factor of $(N!)$.

In this picture, one understands the background values of
$(e^\pm)_{ij}$, used in the gauge functions~\cvrn\/--\cvgh\/, as a
diagonal
background with all eigenvalues equal, that is all $N$ sheets of the
multi-folded surface being non-interacting, flat sheets.
Then clearly, when all matrices commute (matter {\it and} gravity
fields),
one can reasonably expect a cancellation of the anomaly: It should be
proportional to $N$ both in the matter and gravity ghosts sector.
Ideally, in this case, we would set the W--Z type fields $L$, $\bar L$ to
trivial backgrounds.
But if an anomaly was to subsist after  untwisting both
sectors (matrix matter {\sl and} gravity), we would naturally use $L$,
$\bar L$ as W--Z fields, to compensate for the ``residual''
anomaly. We can make them propagate  by adding a sector in the  TQFT
and impose the gauge functions:
\eqn\wz{
(\pa_\z - \be \pa_z) L = \pa_z \be
}
Such constraints  have been used in~\bbg, for the sake of
disentangling left-right asymmetric anomalies. Imposing them as the
gauge functions of a TQFT should
produce a twisted super-Liouville action for the $L, \bar L$
multiplets that
we expect to decouple in the limit of commuting matrices.

Now, the interesting question is  whether one can promote
the independent sheets' reparametrisations into a unified matrix
structure,
in a way such to also ensure anomaly cancellation {\sl away} from
the $IR$
fixed point.
We propose evidence that these ghost systems might arise from
the $\CW_{N+1}$ of $\CW_{\infty}$ partially twisted gravity.

\subsec{$\CW$\/--Gravity proposal}

$\CW_n$--algebras have  $SL(n, {\IR})$ (or $SL_{n}(\IC)$) internal
symmetry~\wwgravity\wwwgravity\wvgravity\surprises.
They can be ghostified, and produce a $2d$--gravity structure with a
duplication of Beltrami differentials and reparametrisation
ghosts. Formally, one can consider their topological extensions,
and reach
a similar ghost action to that of ordinary  topological $2d$--gravity. By
adjusting the value of $n$, one could possibly get a contribution to the
conformal anomaly opposite to that of the matrix superstring. For finite
$n$, technical difficulties arise from the fact that
$\CW_n$--algebras are
not really Lie algebras and need non linear terms to close.
However, these problems disappear for large values of $n$\/:
$\CW_\infty$ is an algebra, for which one can generalize quite
precisely the
notion of Beltrami differentials~\wgravity\/.

Consider $N$ copies of the bosonic string actions in light-cone gauge:
\eqn\nfld{\sum_{i=1}^{N}
\p_{\zb} x^{\mu}_{i} \p_{z} x^{\mu}_{i} + h^{zz}_{i} \p_{z} x^{\mu}_{i}
\p_{z} x^{\mu}_{i} }
Here we allow for a different metric $h^{ab}_{i}$ for each different $i$,
since the Virasoro constraints $T_{zz, i} = 0$ is to be imposed for
each $i$ separately. The way to do so while
preserving the unbroken gauge group $\CS_{N}$ is to form the
$\CW$--like ``currents'' \vafapuzz
$$
\CW_{l} = \sum_{i=1}^{N} T_{zz,i}^{l}
$$
where $l=1, \ldots, N$ and gauge the algebra which is generated by them.

This way, we get $N$ ``ghosts'' $c_{i}$ of increasing conformal
dimension $(-2i, 0)$. In a supersymmetric context these are
accompanied by
the bosonic ghosts of dimensions $({\half} - 2i, 0)$.
Their whole contribution to the conformal anomaly  grows as $N^{2}$
for large $N$, which is a good
indication since we start with $N^{2}$ degrees of freedom in the matter
sector.

\newsec{Further Remarks and Concluding Comments}

\subsec{Six Dimensional  Matrix String Theories}

We could also wonder about a
 similar construction with   the   four dimensional
topological  Yang--Mills theory of~\fourd\/ as a starting point.
Its reduction to two   dimensions,
 already considered in~\cs\/, gives a theory that one can interpret as a
(non-critical) six dimensional {\sl matrix} superstring theory in the
 light-cone gauge (\cf\/ little string theories~\little,\mnmz\/).
With the same chain of arguments as above, the TQFT defined  by
supersymmetrization  would  give its  covariant
 formulation.
All of our formulae are most simply modified to the four dimensional 
case:
 Complex indices should now run from 1 to 2 rather than 3, and the
 octonionic  structure constants should be replaced by those for
quaternions.

The interpretation of this $2d$--theory
goes beyond
the  level of ordinary superstrings since usual perturbative arguments
obviously indicate that it cannot be anomaly free.
However, this might not be so bad a problem:
Although the six-dimensional theory is
perturbatively  anomalous, it could  be  given a sense
  within the non-perturbative version of string theory, where one
can expect an { anomaly inflow} from the bulk, that could take care of 
the
non-conservation of the BRST--current.

\subsec{Matrix Theory}

As an aside, we would like to stress
out that the octonionic or $SU(4)$ instanton equations  also
define a twisted version of the matrix theory of~\bfss\/.
Indeed, after dimensional reduction to zero dimension, one can use the
technique of~\higher\/ and ``oxidize'' to the
 theories in one, two or higher dimensions (the theories in 
\sugino\topmat\
are obtained in this way; see also~\periw).  In zero dimensions, the action is
a sum of squares of commutators. From the positivity of the Casimirs in 
the
topological gauge functions, this action is minimized if and only if each 
one
 of the commutators vanishes. One then has a matrix theory for the Cartan
subalgebra of the group.  This theory has been studied in~\ikkt\ss\mns\/, 
and
the mass deformation in this case is quite useful in analyzing the Witten
index~\mns\/.

With this twisted picture, the
various compactifications (reductions) of M--theory on tori $T^d$ would
then
correspond to lifting-ups (oxidations) of the ``topological matrix 
theory''  to
 dual tori $\tilde{T^d}$.

Moreover, a new feature arises, special to the one dimensional theories: 
When
using the octonionic equations to define some supersymmetric quantum 
mechanics,
one has the possibility of adding a mass term through
\eqn\mass{
D_0 \phi^i + \demi c^i_{jk} [\phi^j, \phi^k] - \mu \phi^i}
where the seven scalars $\phi^{i=1\cdots 7}$\/ are organized in terms of
the fundamental of Spin(7).
This raises opportunities for supersymmetry breaking.

\subsec{Conclusion}

The covariant formulation of the orbifold theory of
DVV, if it can be reached, should produce a covariant formulation of the
second   quantized free string theory.
Once the worldsheet covariance is achieved, the
theory has worldsheet $\Sigma$ with arbitrary
$b_{0}({\Sigma})$ and $b_{1}({\Sigma})$.
Once the formulation is obtained for the $\gs = 0$ limit it
is legitimate to ask for
the extension to non zero $\gs$, i.e.
finite $\ymc$.

In the seek of a covariant formulation, one could have proceeded along 
the
lines  of DVV and  presented a ghost-dressed version of the vertex
operators constructed in~\dvv\/ for ${\IZ}_{2}$ twists (and higher order
twists as well). Adding these operators to the action produces the 
transitions
which change the topology of the worldsheet $\Sigma$. However, since
arbitrary (orientable) topology has already been allowed
by covariance, one gets essentially the same theory. One consequence out
of this redundancy is the possibility to restrict
oneself to genus zero, as was suggested a long time ago
by V.~Knizhnik in~\dima\/. We have not explored this issue in this
paper.

Instead, we have tried to find the matrix
formulation for the covariant theory at finite~$\ymc$, and  have 
suggested
possible formulations of the matrix string theory. Unfortunately, for non
vanishing ${\gs}$, the ones we have worked out so far only make manifest
$SO(8)\times SO(2)$ rather
than the ten--dimensional Poincar\'e symmetry.
Still, we believe that this approach is worth  further
study and could provide hints to the search of the covariant formulation 
of
M--theory itself.

\vskip 1cm

\centerline{\bf Acknowledgements}

We are grateful to A.~Losev,
 A.~Schwarz, S.~Shatashvili, P.~Vanhove   for discussions
and to T.~Banks, M.~Douglas and S.~Shenker for encouragement.
The research of N.~Nekrasov was supported by Harvard Society of
Fellows,
partially by NSF under the  grants PHY-94-07194,
PHY-98-02709, partially by {\cyr RFFI} under grant 96-02-18046 and
partially by grant 96-15-96455 for scientific schools.

\listrefs

\bye